# Explore the Capacity of Near Field Channel using Gaussian Beams

Chenxi Zhu, *Senior member, IEEE*
Wireless Research Laboratory, Lenovo Research, Beijing, China
zhucx1@lenovo.com

*Abstract*—Channel capacity lies at the core of wireless communication, yet determining it typically requires detailed channel information between the transmitter and receiver. For near-field MIMO systems, obtaining the detailed native channel is often difficult or expensive. This paper develops a scheme to approximate the near-field channel in a Gaussian-beam domain. Hermite–Gaussian (HG) modes are used to approximate the channel between a pair of square antenna arrays in a free-space line-of-sight (LOS) environment. We show that HG modes efficiently represent the dominant singular modes of the native channel, enabling accurate channel estimation and capacity computation in the HG beam space. An iterative algorithm is proposed to approach the maximal channel capacity by gradually expanding the beam-space dimension. Simulation results demonstrate that the method converges rapidly and significantly reduces channel-estimation overhead.

## I. INTRODUCTION

Multiple-input multiple-output (MIMO) technology is a cornerstone of modern wireless communications. It can substantially improve user throughput and system capacity under limited spectral resources through beamforming, spatial diversity, and spatial multiplexing. Its importance is expected to grow further in 6G systems with increasingly stringent requirements on capacity, connectivity, and energy efficiency [1][2]. Since its introduction in 4G, MIMO has evolved toward increasingly large antenna arrays. Massive MIMO in 5G marked a transition to large-scale arrays. This trend is expected to intensify in 6G with the adoption of higher frequency bands and even larger array apertures. Under such conditions, the electromagnetic propagation regime changes fundamentally, and many users may lie in the radiative near-field region. In this regime, propagation can no longer be accurately characterized as a superposition of plane waves, and the spherical nature of electromagnetic wave propagation must be explicitly considered. As a result, accurate near-field propagation and channel modeling become essential. This shift introduces new challenges for system design, while also creating opportunities for higher capacity, finer spatial resolution, improved beam focusing, and integrated communication and sensing in next-generation wireless networks [3][4]. A growing body of work has investigated near-field beamforming and precoding, demonstrating their potential to improve spatial multiplexing performance, beam control, and interference management[5][6].

At the core of any wireless communication system lies the question of channel capacity, which is fundamentally determined by the degrees of freedom (DoF) supported by the propagation channel [7]. In principle, the DoF can be obtained from the channel matrix between the transmit (TX) and receive (RX) antennas for arbitrary array geometries, aperture sizes, and propagation conditions. In practice, however, obtaining this channel matrix in near field is often difficult, requiring estimating the channel coefficient for every TX–RX antenna pair under a spherical-wave propagation model before singular value decomposition (SVD) can be performed [8][9][10].

In this paper, we propose a new framework for characterizing the capacity of near-field MIMO channels. Our objective is to determine the maximum channel capacity between the TX and RX arrays under a transmit power constraint, without explicitly requiring the native channel matrix. To overcome the difficulty of obtaining this matrix, we approximate its dominant singular modes using the inherent electromagnetic propagation modes between the TX and RX arrays in the near field. Specifically, the dominant singular modes for a pair of square antenna arrays are represented by a finite set of Hermite–Gaussian beams subject to a prescribed approximation error threshold. The beam-based representation captures not only the transmit and receive field distributions on the apertures, but also the three-dimensional propagation of the electromagnetic field between them. Building on this representation, we develop an algorithm for constructing the channel in a finite-dimensional Hermite–Gaussian beam space, from which the TX and RX spatial filters are obtained by projection onto the corresponding apertures. The channel capacity is then computed iteratively by increasing the beam-space dimension until convergence. The proposed framework yields the maximum channel capacity, the corresponding singular modes (DoF), the associated power allocation, and the TX/RX spatial filters.

## II. Native near field channel for square antenna arrays

We begin by studying the native near-field channel $H$ of the system described in Table I, which consists of a pair of identical square uniform planar antenna arrays facing each other in a free-space line-of-sight (LOS) environment. For simplicity, we assume unidirectional antenna elements. The analysis can be readily extended to cross-polarized antennas, in which case the degrees of freedom are doubled.

Table 1. System Parameters.

| Carrier frequency | 60GHz |
|---|---|
| Bandwidth | 2GHz |
| TX-RX distance | 15m |
| TX/RX antenna array | 27 x 27 elements |
| Antenna spacing | 2cm by 2cm |
| Channel model | Frii's free space model per TX/RX antenna pair |
| Antenna model | 3GPP TS38.901 [11] |
| TX power | -20 dBm |
| RX noise figure | 8 dB |
| Polarization | Unidirectional |
| Power allocation | Water-filling algorithm |
| Modulation and coding scheme | 5G NR modulation and coding up to QAM64 |

The native channel matrix $H$ is computed for every TX/RX antenna pair using the Friis free-space propagation model. Although this computation becomes increasingly expensive for large antenna arrays, it provides a useful reference for the approximation developed later. The singular modes of the native channel, together with the corresponding transmit and receive filters, are obtained through the singular value decomposition (SVD), $H = USV^H$. These singular values are shown in Figure 1.

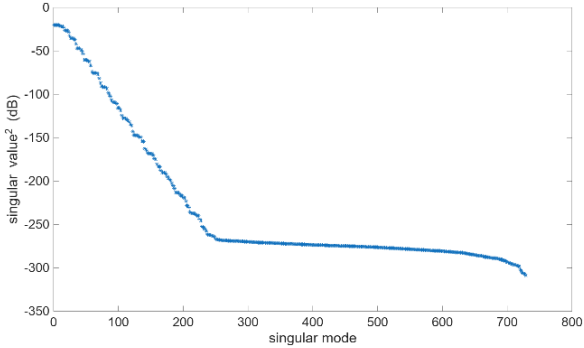

Figure 1. Singular values of the native channel $H$.

Owing to the spherical nature of electromagnetic wave propagation in the near field, reducing the channel rank or identifying the dominant singular modes is much less straightforward than in the far field. In general, the dominant modes cannot be easily identified without computing the full set of singular values. Moreover, the rank alone does not provide meaningful insight into the channel capacity, since many singular modes are too weak to contribute effectively. Under the classic water-filling solution for capacity maximization, transmit power is allocated only to sufficiently strong singular modes, while weak modes receive no power. Therefore, the quantity of practical interest is not the algebraic rank of the channel, but the number of singular modes that are useful for achieving the maximum spectral efficiency under the transmit power constraint. This effective transmission rank depends on both the singular value distribution and the available transmit power.

To gain further insight into the structure of the native channel, we next examine the transmit-side singular vectors. Fig. 2 shows the power profiles of the ten strongest singular modes on the TX array, corresponding to the first ten columns of $V$. Unlike far-field singular modes, which are commonly described by plane waves propagating in specific directions, these near-field modes exhibit localized spatial structures induced by spherical wave propagation. This suggests that a different modal representation than plane-wave is more appropriate for describing the dominant singular modes. Notably, the profiles closely resemble Gaussian beams [12][13]. In [14], we showed that Hermite–Gaussian modes can be transmitted between a pair of square antenna arrays to support parallel multi-stream transmission. Similarly, Laguerre–Gaussian modes can be transmitted between circular apertures in orbital angular momentum systems[15]. Motivated by these observations, in the remainder of this paper we show that Gaussian beams provide an efficient basis for approximating the native channel and achieving its maximal capacity, while also supporting channel estimation, transmission, and reception.

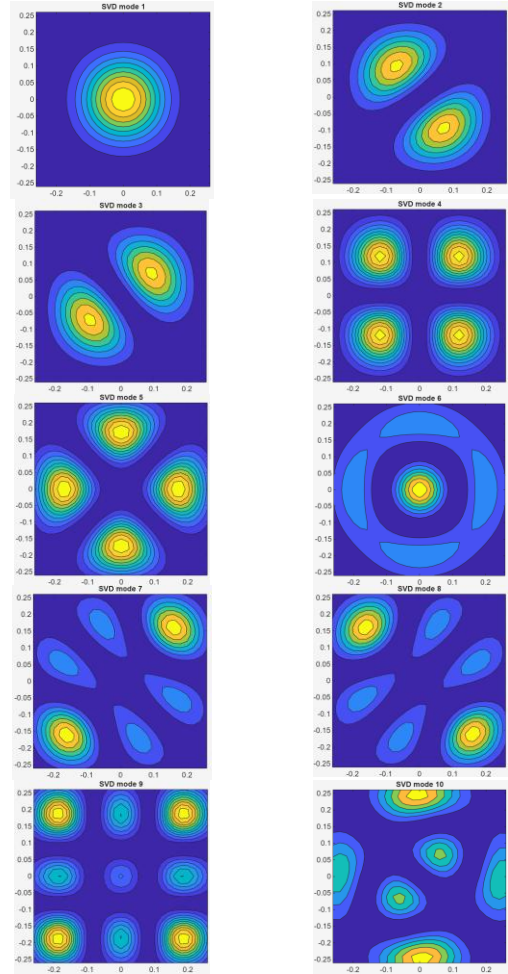

Figure 2. Power profiles of the 10 strongest singular modes of the native channel on the TX array.

## III. GAUSSIAN BEAMS AND HERMITE-GAUSSIAN MODES

In a free-space LOS channel, it is natural to expect that, to maximize throughput, the transmitter should radiate energy primarily toward the receiver. Accordingly, the line connecting the centers of the TX and RX antenna arrays is taken as the dominant propagation direction, i.e., the wave vector $\boldsymbol{k}$, and is chosen as the z-axis. Under this assumption, the signal components that contribute most significantly to channel capacity propagate approximately along $\boldsymbol{k}$. In optics, this regime is referred to as the paraxial regime. For a time-harmonic EM wave propagating in the paraxial regime in free space, the electric field $\boldsymbol{E}$ satisfies the paraxial Helmholtz equation

$$\left(\nabla^2 + \frac{\omega^2}{c^2}\right)\boldsymbol{E} = (\nabla^2 + k^2)\boldsymbol{E} = 0 \quad (1)$$

where $k = \frac{2\pi}{\lambda}$ and where $\nabla_T^2$ denotes the Laplacian operator in the transverse plane. Solving (1) in different coordinate systems yields different families of paraxial solutions, commonly referred to as Gaussian beams. Typical examples include Hermite–Gaussian (HG) beams, Laguerre–Gaussian (LG) beams, and Ince–Gaussian (IG) beams [12][13][16], which arise from solutions in Cartesian, cylindrical, and elliptical coordinates, respectively. Each family forms a complete basis for paraxial fields, and transformations between different bases are possible[13]. Among these beam families, HG beams are particularly attractive because their rectangular spatial structure is well matched to square antenna apertures. This makes them a natural candidate for representing the singular modes of the native channel. As will be shown later, selecting an appropriate Gaussian-beam basis can significantly reduce the complexity of channel estimation and transmission mode computation.

Assume the focal plane is at $z = 0$. In the Cartesian coordinate $\nabla_T^2 = \frac{\partial^2}{\partial x^2} + \frac{\partial^2}{\partial y^2}$, solving (1) gives a set of orthonormal solutions, each characterized by a pair of integer mode number $(l, m)$:

$$HG_{l,m}(x, y, z) = \sqrt{\frac{1}{2^{l+m-1}\pi l! m!}} \frac{1}{w(z)} H_l\left(\frac{\sqrt{2}x}{w(z)}\right) H_m\left(\frac{\sqrt{2}y}{w(z)}\right) \exp\left(-\frac{x^2+y^2}{w^2(z)}\right) \cdot$$
$$\exp\left(-i\frac{k(x^2+y^2)}{2R(z)}\right) \exp\left(i\psi_{l,m}(z)\right) \exp(-ikz) \quad (2)$$

Each mode $HG_{l,m}(x, y, z)$ is normalized and represents an orthogonal beam traveling along the z-axis. $H_l(\cdot)$ is the Hermite polynomial (the physicist type) of order $l$. Hermite polynomials of different orders are mutually orthogonal on the infinite XY plane with respect to the weight function $e^{-x^2}$:

$$\int_{-\infty}^{\infty} H_m(x) H_n(x) e^{-x^2} dx = \sqrt{\pi}\, 2^n n!\, \delta_{m,n} \quad (3)$$

where $\delta_{m,n}$ is the Kronecker delta function. $H_l(x)$ and $H_m(y)$ determines the beam profiles in the x and y directions respectively. $Z_{Ray} = \frac{\pi w_0^2}{\lambda}$ is the Rayleigh distance, and the radius of the curvature at the wavefront is $R(z) = \frac{z^2 + Z_{Ray}^2}{z}$. $\psi_{l,m}(z) = (1 + l + m)\arctan\left(\frac{z}{Z_{Ray}}\right)$ is Gouy's phase. The beam radius at $z$ is $w(z) = w_0\sqrt{1 + \frac{z^2}{Z_{Ray}^2}}$. Different modes remain mutually orthogonal during propagation. Fig. 3 shows the power profiles of the first few HG modes on the TX antenna array.

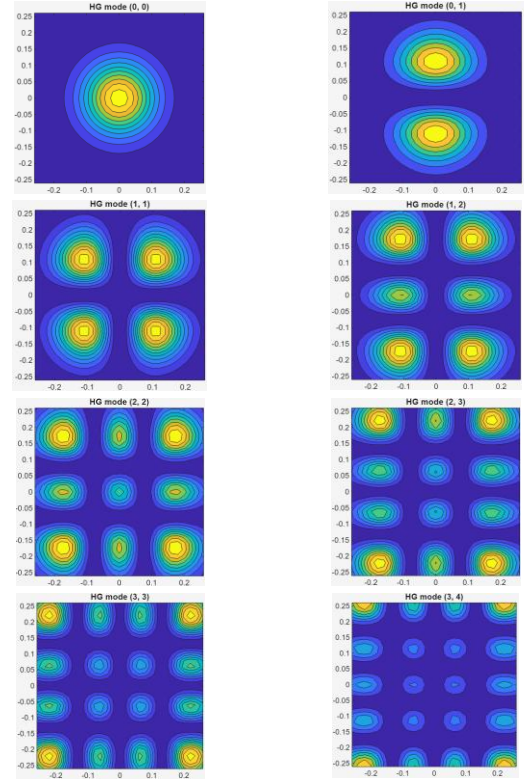

Figure 3. Beam profiles of some HG modes. The profiles of $(l, m)$ and $(m, l)$ are identical after 90° rotation.

A comparison between Fig. 2 and Fig. 3 reveals a clear similarity between the dominant singular-mode profiles of the native channel and lower order HG mode patterns. This observation suggests that the singular modes can be efficiently represented using a finite set of HG modes. In principle, the Hermite–Gaussian modes form an orthonormal basis for paraxial fields over an infinite transverse plane. In the present problem, however, transmission and reception take place over finite TX and RX apertures, while the standard HG modes extend infinitely in the xy-plane. On the antenna apertures the HG modes are no longer orthogonal. The practical efficiency of the representation depends on how much of each mode is captured by the finite antenna arrays. This effect is illustrated in Fig. 4, which shows the normalized power captured by different HG modes as a function of the normalized antenna size. For a given mode order, the captured power increases as the antenna aperture becomes larger. Conversely, higher-order modes retain less power for a fixed aperture size. These observations indicate that the choice of HG beam parameters is critical for finite-aperture implementations. A smaller beam radius at the antenna plane is desirable for two related reasons. First, tighter beam confinement allows more useful HG modes to be accommodated within the limited TX and RX apertures, thereby increasing the number of spatial degrees of freedom available for parallel transmission. Second, because practical antenna arrays truncate the ideal infinite-plane HG modes, excessive beam spreading causes greater power loss outside the aperture and degrades the orthogonality of the truncated modes. By reducing the beam radius at the antenna plane, this truncation effect is mitigated, and the finite-aperture modes remain closer to orthogonal. As a result, a more compact beam footprint improves both modal efficiency and the effective number of useful transmission modes.

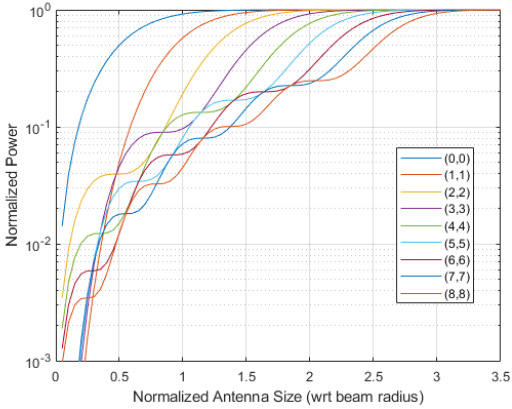

Figure 4. Normalized power captured by different HG modes versus normalized antenna size.

The spatial extent of an HG beam is determined by its beam radius $w(z)$, which depends on both the propagation distance z from the focal plane and the beam waist $w_0$. Since the Rayleigh distance $Z_{Ray}$ is itself a function of $w_0$, the beam radius at the array plane can be optimized through an appropriate choice of beam parameters. When the TX and RX arrays are identical and separated by a distance $D_{TR}$, it is natural to place them symmetrically with respect to the focal plane, i.e., $-z_T = z_R = 0.5 D_{TR}$, so that the beam radii at the TX and RX planes are equal. Under this symmetric configuration, the beam radius at the RX plane is

$$w_{TX}(z_T) = w_0 \sqrt{1 + \frac{z_T^2}{Z_{Ray}^2}} = w_0 \sqrt{1 + \frac{z_T^2 \lambda^2}{\pi^2 w_0^4}} \quad (4)$$

which is minimized when

$$w_0^* = \sqrt{\frac{\lambda z_R}{\pi}}, \quad (5)$$

yielding the optimal beam radius at the TX and RX planes,

$$w_{TX}^*(z_T) = w_{RX}^*(z_R) = \sqrt{\frac{\lambda D_{TR}}{\pi}} = \sqrt{2} w_0^*. \quad (6)$$

Therefore, the optimal symmetric configuration places both the TX and RX arrays one Rayleigh distance away from the focal plane, i.e., $Z_{Ray}^* = z_R = -z_T$. With the optimal beam parameters determined, we next evaluate how many HG modes are required to represent the singular modes of the native channel. To this end, we decompose the singular modes of the native channel onto the HG basis and examine the resulting residual error. Due to their similarities, the singular-modes are expected to be well represented by the HG modes.

Assume that the TX and RX array apertures are defined by $|x|, |y| \leq a$ in the xy-plane. Then, the electric field corresponding to the $k$-th singular mode on the TX array, denoted by $E_{SVD}^k$ and given by the $k$-th column of $V$, can be expanded as a linear combination of HG modes:

$$E_{SVD}^k(x, y, z_T) = \sum_{l,m} E_{l,m}^k HG_{l,m}(x, y, z_T), |x|, |y| \leq a. \quad (7)$$

Let the TX antenna elements be indexed by $-N \leq i, j \leq N$. Evaluating the field at the discrete antenna locations $(x_i, y_j)$ gives

$$E_{SVD}^k(x_i, y_j, z_T) = \sum_{(l,m)} E_{l,m}^k HG_{l,m}(x_i, y_j z_T), -N \leq i, j \leq N. \quad (8)$$

Because the antenna aperture is finite, the HG modes are no longer strictly orthogonal over the sampled TX array. Therefore, they must first be re-orthonormalized over the aperture before the expansion coefficients are computed. After this re-orthonormalization, the coefficient corresponding to mode $(l, m)$ can be obtained as

$$E_{l,m}^k = \sum_{-N \leq i, j \leq N} E_{SVD}^k(x_i, y_j, z_T) HG_{l,m}^*(x_i, y_j, z_T). \quad (9)$$

Let $L_{max}$ denote the maximum mode index in either the x- or y-direction, i.e., $0 \leq l, m \leq L_{max}$. The corresponding HG beam space then contains $(L_{max} + 1)^2$ modes. Within this truncated beam space, we define the residual representation error for the $k$-th singular mode as

$$err_{L_{max}}^k = \frac{\sum_{-N_t \leq i,j \leq N_t} \left| E_{SVD}^k(x_i, y_j z_T) - \sum_{0 \leq l, m \leq L_{max}} E_{l,m}^k HG_{l,m}(x_i, y_j z_T) \right|^2}{\sum_{-N_t \leq i,j \leq N_t} \left| E_{SVD}^k(x_i, y_j z_T) \right|^2}. \quad (10)$$

Fig. 5 shows the residual error for the singular modes obtained using truncated HG spaces with different values of $L_{max}$. For a given $L_{max}$, the residual error remains small for the first $(L_{max} + 1)^2$ singular modes, with the effect being particularly pronounced for $L_{max} \leq 3$. This indicates that the dominant singular modes of the native channel are primarily composed of low-order HG modes, whereas the contribution of higher-order HG modes can be ignored with little loss of accuracy. This analysis applies to the RX array as well as the TX array.

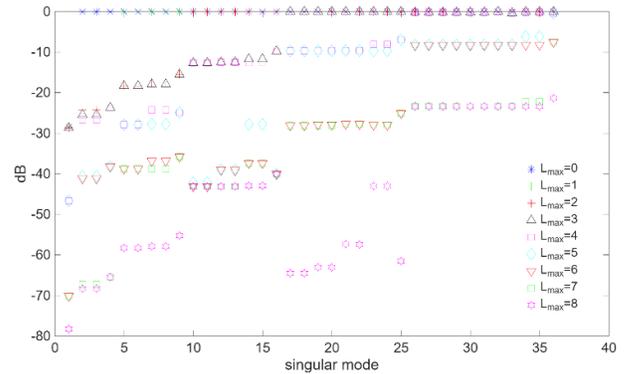

Figure 5. Residual error of the singular values after representation with HG modes $0 \leq l, m \leq L_{max}$.

Since capacity-achieving transmission under the TX power constraint relies primarily on the strongest singular modes, this result further suggests that only a limited number of HG modes are needed. In other words, the HG beam space required for capacity characterization can be kept relatively small.

IV. Explore channel capacity in the HG beam space

Motivated by the efficient representation of the dominant singular modes, we now construct the channel directly in the HG beam domain. To this end, we construct a channel matrix $H_{HG}$ whose input and output dimensions correspond to transmit and receive HG modes, respectively. In practice, $H_{HG}$ can be estimated by transmitting and receiving reference signals in the HG domain. Specifically, the channel coefficient between the TX mode $(l,m)$ and the RX mode $(n,p)$ can be estimated by transmitting a reference signal using $HG_{l,m}(:,:,z_T)$ as the TX spatial filter and applying $HG_{n,p}^*(:,:,z_R)$ as the RX spatial filter to the received signal. Standard channel estimation methods, such as least squares (LS) or minimum mean-square error (MMSE) estimation, can then be used to obtain the channel coefficient between the corresponding TX and RX beam modes. Repeating this procedure for all selected HG mode pairs yields the HG-domain channel matrix $H_{HG}$. Applying singular value decomposition $H_{HG} = U_{HG} S_{HG} V_{HG}^H$ gives the singular modes in the HG beam space. Fig. 6 compares the singular values of $H_{HG}$ with those of the native channel $H$ under ideal channel estimation. The two sets of singular values are in close agreement, particularly for the dominant modes. This result confirms that the selected HG beam set provides an accurate representation of the system and that the channel in the HG beam space is effectively equivalent to the native channel for the modes relevant to transmission.

An additional advantage of the HG-domain representation is reduction in channel estimation overhead. Let $H_{HG}^{L_{max}}$ denote the HG-domain channel matrix constructed using $(1 + L_{max})^2$ modes with $0 \leq l, m \leq L_{max}$. Estimating $H_{HG}^{L_{max}}$ requires transmitting reference signals only over these $(1 + L_{max})^2$ TX modes, rather than over every physical transmit antenna element. Each transmitted reference signal is then observed through the selected receive HG modes to recover the corresponding channel coefficients. As a result, channel estimation is performed in a much lower-dimensional domain. As an example, only 81 HG-mode reference signals are required in Fig. 6, compared with 729 reference signals in the antenna domain, corresponding to an overhead reduction of 89%. Moreover, a reference signal transmitted in an HG mode is distributed across the array aperture, which can provide a higher effective transmitted energy than exciting individual antennas separately. Higher transmission power improves channel estimation performance at the receiver. Since the dominant singular modes are captured accurately in the truncated HG space, it is sufficient to choose $L_{max}$ only large enough to meet the desired approximation accuracy.

Based on the preceding results, we develop an algorithm to estimate the channel capacity $C_{max}$ using a limited set of Hermite–Gaussian (HG) modes. The algoritm incrementally increases the maximum mode index $L_{max}$ while monitoring the convergence of the achievable capacity obtained from the $(1 + L_{max})^2$ HG modes. Let $C_{HG}(H_{HG}^i)$ denote the channel capacity computed from the estimated HG-domain channel matrix $H_{HG}^i$, using the standard water-filling algorithm under the transmit power constraint.

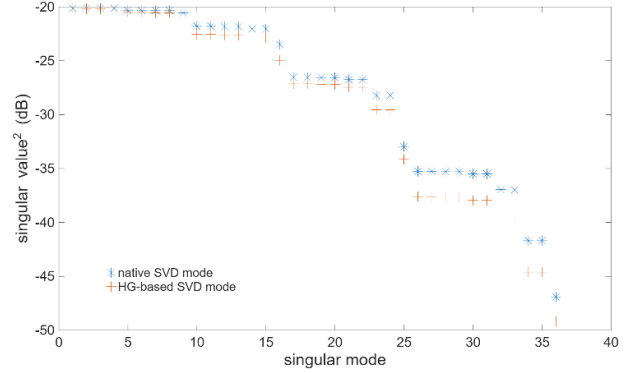

Figure 6. Comparison of the singular values of the native channel $H$ and the HG-domain channel $H_{HG}^8$.

**HG-Mode-Based Channel Capacity Calculation Algorithm**
**Input:** Tolerance $\varepsilon$, transmit power constraint.
**Output:** Estimated capacity $C_{max}$, maximum mode index $L_{max}$.
1. **Initialization:**
   Estimate $H_{HG}^0$ and $H_{HG}^1$ using reference signals transmitted in HG modes (0,0),(0,1),(1,0),(1,1).
   Set $i = 1$.
2. **Repeat:**
   Compute $C_{HG}(H_{HG}^i)$ via water-filling.
   **while** $(|C_{HG}(H_{HG}^i) - C_{HG}(H_{HG}^{i-1})| > \varepsilon)$
   do
    $i \leftarrow i + 1$;
    Transmit reference signals in modes $(0,i),(1,i),\ldots,(i,i),(i,i-1),\ldots,(i,0)(0,i), (1,i)$,
    Update channel estimate $H_{HG}^i$
    Compute $C_{HG}(H_{HG}^i)$
   **end while**
3. **Output:**
   $L_{max} = i, \ C_{max} = C_{HG}\left(H_{HG}^{L_{max}}\right)$.

Fig. 7 illustrates the computed channel capacity (spectral efficiency) as a function of $L_{max}$. The estimated capacity rapidly converges to the true capacity (shown as a red line) as $L_{max}$ increases. At $L_{max}$=8,9,10,11, the capacity is within 5%, 3%, 1.4%, and 0.59% of the true capacity, respectively. At $L_{max} = 12$, the selected modes, corresponding power allocation, and resulting spectral efficiency match those obtained from the native channel matrix $H$. No further variation is observed beyond this point. Notably, all computations are performed entirely in the HG beam domain, without requiring explicit knowledge of $H$. At convergence, $L_{max}$ represents the highest mode index needed to satisfy the tolerance $\varepsilon$, and a total of $(1 + L_{max})^2$ HG modes are used. These modes effectively represent the corresponding number of dominant singular modes of the original channel, thereby achieving near-optimal capacity.

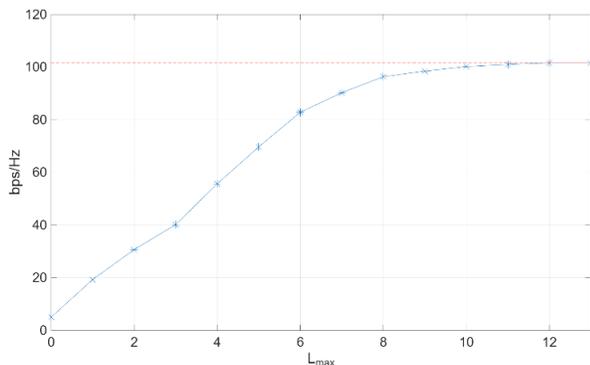

Figure 7. Spectrum efficiency as a function of the maximal mode number $L_{max}$. The red line is the maximal spectrum efficiency (true capacity) calculated using the native channel matrix $\boldsymbol{H}$.

## V. OTHER GAUSSIAN BEAM FAMILIES

Because all Gaussian beam families form complete bases, any of them can be employed for modal representation. The selection primarily depends on how closely their beam profiles resemble the singular modes of the native channel for a given antenna array geometry and size. Hermite–Gaussian (HG) modes, characterized by rectangular symmetry, are well suited to square or rectangular antenna arrays. In contrast, Laguerre–Gaussian (LG) modes exhibit circular symmetry and are therefore more compatible with circular apertures. This correspondence explains why orbital angular momentum (OAM) systems employing circular arrays are commonly described using LG modes. Another member of the Gaussian beam family is the Ince–Gaussian (IG) modes[16]. Notably, certain singular modes of square antenna arrays exhibit strong resemblance to specific IG modes. As illustrated in Fig. 8, the eighth singular mode of the native channel closely resembles the IG mode (4,4) with even parity. This striking similarity suggests that the Ince–Gaussian beam family may provide comparable performance in such configurations.

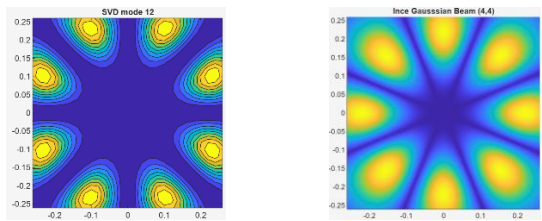

Figure 8. Comparison of $8^{th}$ singular mode of the native channel with Ince-Gaussian mode (4,4), ellipticity=0.02, parity=even.

## VI. CONCLUSION AND FUTURE WORK

This paper presents a scheme for achieving the capacity of line-of-sight (LOS) near-field channels using Gaussian beam representations. Specifically, we demonstrate that the Hermite–Gaussian (HG) beam family provides an efficient basis for representing the singular modes of the native channel between square antenna arrays, enabling near-capacity performance. By performing channel estimation and capacity computation directly in the HG beam domain, the proposed approach significantly reduces the number of reference signals required. An iterative algorithm is further developed to progressively increase the number of HG modes used for capacity evaluation based on channel estimates obtained in a reduced beam space. Numerical results show that the algorithm converges rapidly to the true channel capacity.

The present study assumes an ideal free-space channel model. As future work, more realistic propagation conditions, including scattering effects, will be investigated. In addition, the current formulation assumes known transmitter–receiver separation and perfect alignment between antenna arrays. In practical systems, these parameters may be unknown or imperfect and must be jointly estimated as part of the channel. This motivates the development of more robust channel estimation techniques.